\documentclass[prl,twocolumn,showpacs,preprintnumbers]{revtex4}
\usepackage{graphicx}
\usepackage{amsmath}
\usepackage{amssymb}
\def\s{\sigma}

\def\ua{\uparrow}
\def\da{\downarrow}
\def\dtau{\varDelta \tau}
\def\Vec#1{\mbox{\boldmath $#1$}}
\begin{document}

\preprint{submitted to Phys. Rev. B}

\title{A new numerical algorithm for the double-orbital Hubbard model\\
--- Hund-coupled pairing symmetry in the doped case}

\author{Shiro Sakai, Ryotaro Arita, Hideo Aoki}

\affiliation{Department of Physics, University of Tokyo, Hongo,
Tokyo 113-0033, Japan}

\date{\today}

\begin{abstract}
In order to numerically 
study electron correlation effects in multi-orbital systems, 
we propose a new type of discrete 
transformation
for the exchange (Hund's coupling) and pair-hopping interactions 
to be used in the dynamical mean field theory + quantum Monte Carlo method. 
The transformation, which is real and exact, turns out to suppress 
the sign problem in a wide parameter region including 
non-half-filled bands.  This enables us to obtain 
the dominant pairing symmetry in the double-orbital Hubbard model, 
which shows that the spin-triplet, orbital-antisymmetric 
pairing that exploits Hund's coupling is stable 
in a wide region of the band filling.
\end{abstract}
\pacs{71.10.Fd,74.20.Rp,71.30.+h}
\maketitle

In the physics of correlated electron systems, as highlighted 
by the high-$T_c$ superconductivity, 
the prototypical transition-metal oxides have arrested attention
because of a rich variety of physical properties, among which are 
anisotropic pairing symmetries in the cuprates and in $\rm Sr_2RuO_4$, 
the colossal magnetoresistance in manganites, and a complex phase
diagram for $\rm La_{1-x}Sr_xMnO_3$.\cite{tn00}
Since the $d-$orbitals are relevant, 
we are actually talking about multi-orbital systems, and, 
while some of the phenomena should be generically captured within single-band 
models, understanding of the effect of electron correlation in 
multi-orbital system remains a fundamental problem.\cite{roth66} 

Namely, the orbital degrees of freedom should couple to other 
degrees of freedom 
such as charge, spin, and lattice distortion, and we can expect even 
richer physical properties. Indeed, 
the colossal magnetoresistance results from Hund's coupling
and the complex phase diagrams of Mn and Co compounds 
are considered to arise from a competition between Hund's 
coupling and the Jahn-Teller distortion.
In perovskite-type oxides, the crystal field 
splits the five $3d-$orbitals  into three-fold degenerate ($t_{2g}$) and 
two-fold degenerate ($e_g$) levels.  
When the degenerate levels are not fully filled, the degeneracy 
may be lifted by the Jahn-Teller effect, where the system is effectively 
mapped to a single-orbital model when the splitting is large enough.  
We take here the small splitting limit to concentrate on 
the physics specifically caused by the orbital degrees of freedom.  

Both numerical and analytic methods have been developed to study 
correlated electron systems.  The methods should preferably be 
non-perturbative if one wants to examine the effect specific to the 
electron correlation such as Mott's metal-insulator transition.  
The dynamical mean field theory (DMFT)\cite{gkkr}, which can fully include 
temporal fluctuations while spatial fluctuations are neglected,
first succeeded in describing this transition both from metallic and 
insulating sides.  
In this method a lattice system (such as the Hubbard model) is 
mapped to an impurity model, which becomes 
an exact mapping in the limit of infinite spatial dimension.  
A standard procedure is to solve the impurity problem with the 
(auxiliary-field) quantum  Monte Carlo(QMC) method\cite{h85,j92}, 
which involves no approximations except for the Trotter decomposition.  
So the DMFT+QMC method should be a desirable candidate for the multi-orbital 
cases.  Unfortunately, it is difficult to extend 
the QMC method to multi-orbitals:  First, it is impossible to express 
those (exchange and pair-hopping) interactions that are specific to 
multi-orbital cases in terms of the usual 
auxiliary fields.
Second, even when we can accomplish this, the negative sign problem, 
a notorious problem in QMC calculations, is usually difficult 
to avoid for multi-orbitals.  

This has motivated us to propose here a new auxiliary-field 
transformation that is applicable to Hund's and pair-transfer terms.  
The transformation, which is real and discrete, turns out to suppress 
the sign problem in a wide parameter region including non-half-filled bands.
This enables us to examine the role of Hund's coupling and pair-transfer 
in a double-orbital model.
One of the most intriguing questions for correlated electrons on multi-orbits 
is what should be the symmetry of the superconducting pairing that arises 
from the electron-electron interaction.
We have examined the dominant pairing symmetry in the double-orbital 
Hubbard model with the DMFT+QMC method.  
The result reveals that the spin-triplet $\otimes$ orbital-antisymmetric 
$\otimes$ even-frequency pairing, which exploits Hund's coupling, 
is stable in a wide region of the band filling.

So we take the two-fold degenerate Hubbard 
model\cite{oles83} with a Hamiltonian, 
 \begin{eqnarray}\label{eq:hamiltonian}
  &H&= H_0+H_1+H_2,\\
  &H_0& = -t\sum_{ij\s} \sum_{m}^{1,2} c_{im\s}^\dagger c_{jm\s} 
  -\mu\sum_{im\s}n_{im\s},\nonumber\\
  &H_1& = U\! \sum_{i,m} n_{im\uparrow}n_{im\downarrow}
         +U'\sum_{i\s} n_{i1\s} n_{i2-\s}\nonumber\\
      &&  +(U'-J)\sum_{i\s} n_{i1\s} n_{i2\s},\nonumber
 \end{eqnarray}
 \begin{eqnarray}
  &H_2&\! =\! J\!\!\sum_{i,m\ne m'}(
          c_{im\ua}^\dagger c_{im'\da}^\dagger c_{im\da} c_{im'\ua}
         +c_{im\ua}^\dagger c_{im\da}^\dagger  c_{im'\da}c_{im'\ua}),\nonumber
 \end{eqnarray}
Here $c_{im\s}^\dagger$ creates an electron of spin $\s$ in the orbital 
$m(=1,2)$ at site $i$, and 
$n_{im\s}\equiv c_{im\s}^\dagger c_{im\s}$. 
We only consider the nearest-neighbor hopping between the similar orbitals, 
electron-electron interactions are assumed to be intra-atomic with 
the intra-(inter-)orbital Coulomb interaction denoted as $U(U')$, 
while the exchange and pair-hopping interactions as $J$.  
The Hamiltonian is rotationally invariant not only in spin, but also 
in real space if we fulfill the condition $U\!=\!U'\!+\!2J$ 
(as is the case with $d$-orbitals).  
We have divided the interaction into 
the density-density interactions $H_1$ and the exchange and pair-hopping 
interactions $H_2$.

The DMFT+QMC method has been used by many authors for the single-orbital 
Hubbard model.  
For the usual on-site Hubbard interaction, a decoupling is done 
with the discrete Hubbard-Stratonovich(HS) transformation\cite{h83},
\begin{align}\label{eq:H-S}
 e^{-a [n_{\uparrow}n_{\downarrow}
         -\frac{1}{2}(n_{\uparrow}+n_{\downarrow})] }
 \!=\! \left\{\begin{array}{l}
 \frac{1}{2}\sum_s^{\pm 1}e^{\lambda s(n_{\uparrow}-n_{\downarrow})}
 (a\geq 0) \\
 \frac{1}{2}\sum_s^{\pm 1}e^{\lambda s(n_{\uparrow}+n_{\downarrow}-1)
 +\frac{a}{2}}  (a<0) \end{array} \right.\!,
\end{align}
with $\lambda \equiv \log (e^{\frac{\left| a\right|}{2}}
                    +\sqrt{e^{\left| a \right|}-1})$, 
which transforms the two-body interaction 
into one-body interactions summed over an auxiliary field $s$.
Applying this to each interaction term on the discretized
imaginary time, we can decompose the partition function $Z$ of the 
many-body system into a sum of the partition functions $Z_{\{s_i\}}$ of 
one-body systems as 
$
Z = \sum_{\{s_i\}} Z_{\{s_i\}}.
$
The QMC samples the single-particle systems according to the weight 
$Z_{\{s_i\}}$.  The negative sign refers to the fact that the weight is not 
positive-definite.  The sign problem does not occur, as far as 
the DMFT is concerned, in the single-orbital Hubbard model because 
the impurity problem lacks the electron hopping terms.

While the auxiliary-field QMC method has been applied to some multi-orbital 
Hubbard models by neglecting the terms other than the density-density 
interactions ($H_1$ in eq.(\ref{eq:hamiltonian})), 
the QMC algorithm becomes a challenging problem 
for the exchange and pair-hopping interactions ($H_2$): 
the HS transformation (\ref{eq:H-S}) is obviously inapplicable to these 
terms.  While a decoupling, 
${\rm exp}(J \dtau c_{1}^\dagger c_{2}c_{3}^\dagger c_{4}) 
= (1/2)\sum_s^{\pm 1}{\rm exp}[s\sqrt{J\dtau} (c_{1}^\dagger c_{2}
                                   -c_{3}^\dagger c_{4})],$
is possible after breaking $e^{-\dtau H_2}$ into a product of 
exponentials, it leads to a serious sign problem\cite{hv98}.
Another attempt by Motome and Imada\cite{mi97} decouples 
$H_2$ with imaginary auxiliary fields to implement a QMC, 
but an electron-hole symmetry has to be assumed to avoid 
phase cancellation of the weights which become complex due to the 
imaginary auxiliary fields.  
Since the assumption dictates the half-filled band and $U\!=\!U'$ (that 
violates $U\!=\!U'\!+\!2J$), investigation of non-half-filled bands 
and/or the rotationally symmetric case is difficult with this method.
   
We propose here a new type of discrete 
transformation for the exchange and pair-hopping interaction terms, 
 \begin{align}\label{eq:real}
  e^{-\dtau H_2} &=&
  \frac{1}{2}\sum_{r=\pm 1} e^{\lambda r (f_{\ua}-f_{\da})}
  e^{a(N_{\ua} + N_{\da})+bN_{\ua}N_{\da}},
  \end{align}
where
 \begin{eqnarray*}
  \lambda &\equiv& \frac{1}{2}\log(e^{2J\dtau }+\sqrt{e^{4J\dtau}-1}),\\
  a &\equiv& -\log(\cosh(\lambda)),\quad b \equiv \log(\cosh(J\dtau)),\\
  f_{\s} &\equiv& c_{1\s}^\dagger c_{2\s} + c_{2\s}^\dagger c_{1\s},\quad
  N_{\s} \equiv n_{1\s} + n_{2\s} -2n_{1\s} n_{2\s}.
 \end{eqnarray*}
This transformation is exact with the auxiliary field ($r$) being real 
and all the operators being hermitian.
Although a term $N_{\ua}N_{\da}$ which is forth order in $n$ appears 
on the right hand side, we can apply the usual 
HS transformation to this term due to a property
 $  N_{\s}^2 \!=\! N_{\s}$.
The resulting terms in the form $nn$ can be combined with the Coulomb terms,
which are transformed with eq.(\ref{eq:H-S}), so that we need in total only two 
auxiliary-fields for $H_2$.
Furthermore the interaction parameters $U,U',J$ can be varied independently, 
which means that we can treat the rotational symmetric 
cases of $U\!=\!U'\!+\!2J$. 
We note that recently Han\cite{han04} proposed another type of 
HS transformation for the Hamiltonian (\ref{eq:hamiltonian}).
He adopts a continuous HS transformation in combination with the discrete one 
(\ref{eq:H-S}), 
and avoided the sign problem in a wide parameter region at half-filling. 
In contrast all the transformation are discrete in the present method.

 \begin{figure}[t]
 \includegraphics[width=7cm,height=5.5cm]{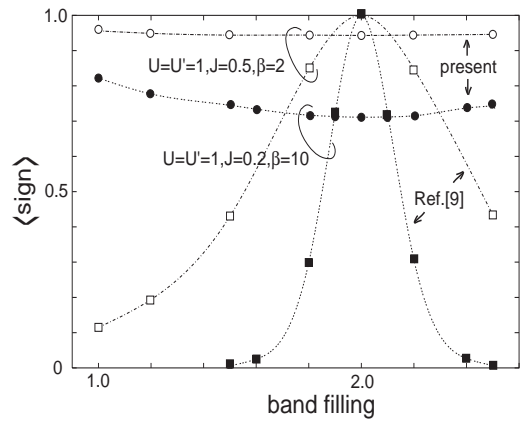}
 \caption{The average sign plotted as a function of  
          band filling at 
$\beta\!\equiv 1/T =\!2$ 
for $U\!=\!U'\!=1,~J\!=\!0.5$ and at $\beta\!=\!10$ for 
$U\!=\!U'\!=\!1,~J\!=\!0.2$ calculated with the present algorithm 
          (circles) and with the one due to \cite{mi97} (squares). 
In this particular plot we have set $U\!=\!U'$ to facilitate comparison with Motome and Imada's method.\cite{commentelhole}  
          Curves are guides to the eye here and in subsequent figures.}
 \label{fig:ndep}
 \end{figure}

 \begin{figure}[t]
 \centering{
 \includegraphics[width=7cm,height=5.5cm]{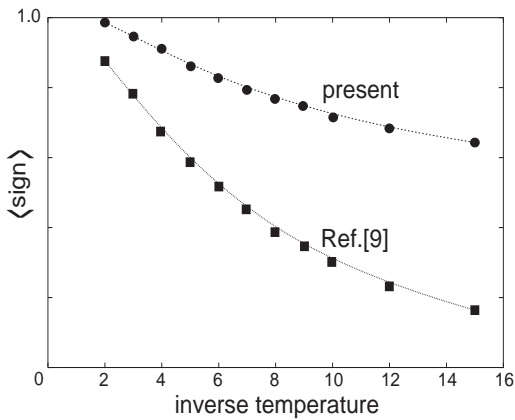}
 \caption{The average sign plotted as a function of 
          inverse temperature for $n\!=\!1.8,~U\!=\!U'\!=\!1,~J\!=\!0.2$ 
          calculated with the present algorithm 
          (circles) and with the one due to \cite{mi97} (squares). 
          }
 \label{fig:bx18j02}}
 \end{figure}

One valuable property of the transformation (\ref{eq:real})
is that negative weights are indeed reduced significantly.
An unexpected asset is that the suppression of the sign problem 
is almost independent of the band filling.  
This is shown in Fig.\ref{fig:ndep}, where the average sign 
$\langle {\rm sign} \rangle \equiv \sum_{\{s_i\}}Z_{\{s_l\}}/\sum_{\{s_i\}}|Z_{\{s_l\}}|$, is plotted as a function of the band 
filling for two sets of values of $\beta\!\equiv 1/T$ and $J$ 
for the two-orbital Hubbard model on an 
infinite dimensional hypercubic lattice, whose density of states is 
Gaussian with the band width $W\!=\!2$.  
We can see that the sign problem is avoided almost independently of 
the band filling.  
Figure \ref{fig:bx18j02} depicts the temperature dependence, which shows that 
$\langle {\rm sign} \rangle$ decreases mildly as the temperature lowers, so 
we can go to lower temperatures with the present transformation. 
One reason why the transformation (\ref{eq:real}) reduces the amount of 
negative weights is that it has a single 
auxiliary field ($r$ in eq.(\ref{eq:real})) as the source of the negative sign;
the other auxiliary fields, related to the density-density interactions, are 
irrelevant to the negative weights.

To test how the present method works, 
we have applied it to the calculation of the superconducting 
susceptibility for the Hamiltonian (\ref{eq:hamiltonian}).  
Since off-site, anisotropic pairings (such as $p$ and $d$-waves) 
cannot be treated within the DMFT,
we confined ourselves to $s$-wave pairing.  
Even within that channel, various pairing symmetries 
are possible in multi-orbital systems, since the total 
symmetry now consists of spin $\otimes$ orbital $\otimes$ frequency, so  

\begin{tabular}{c|ccc}
      & spin & orbital & frequency \\ \hline
 1SE&  singlet & symmetric & even \\
 3AE&  triplet & antisymmetric & even \\
 1AO&  singlet & antisymmetric & odd \\
 3SO&  triplet & symmetric & odd \\
\end{tabular}

\noindent are the possibilities.
The pairs that are formed 
across different orbitals are especially interesting.  
The orbital-symmetric pairs are
\begin{align}\label{eq:basis}
 {\rm S^a}: \quad c_{1\ua}c_{1\da}+c_{2\ua}c_{2\da},\nonumber\\
 {\rm S^b}: \quad c_{1\ua}c_{1\da}-c_{2\ua}c_{2\da},\\
 {\rm S^c}: \quad c_{1\ua}c_{2\da}+c_{2\ua}c_{1\da},\nonumber
\end{align}
where $c_{1\ua}c_{1\da}$ and $c_{2\ua}c_{2\da}$ are combined into 
bonding and antibonding states due to the pair-hopping term, 
while the orbital-antisymmetric pair is 
\begin{eqnarray*}
 {\rm A}: \quad c_{1\ua}c_{2\da}-c_{2\ua}c_{1\da}.
\end{eqnarray*}
Here we take, without the loss of generality, the $S_z=0$ channel for 
spin triplets, where $S_z$ is the $z$-component of the Cooperon spin.

So the interest here is which pairing symmetry is favored in the 
double-orbital system, especially in the presence of Hund's coupling.  
For the single-orbital case the sites are mostly singly-occupied by electrons 
around the half-filling, where the pairing (usually in $d$-wave channel) 
occurs. The question is what would be the corresponding picture for the 
double-orbital case around half-filling.  
So we have calculated 
the pairing susceptibility $P$, which is related to 
the two-particle Green function\cite{jm01},
 \begin{eqnarray}\label{eq:chi}
   \lefteqn{\chi_{ll',mm'}(i\omega_n,i\omega_n') }\nonumber\\ 
   &\equiv&\!\!\int_0^\beta\!\int_0^\beta\!\int_0^\beta\!\int_0^\beta\! 
   d\tau_1 d\tau_2 d\tau_3 d\tau_4
   e^{i[\omega_n(\tau_1-\tau_2)+\omega_n'(\tau_3-\tau_4)]}\nonumber\\
   &\times&\!\!\sum_{\Vec{kk}'}\langle{\rm T}_\tau 
        c_{\Vec{k}l'\ua}(\tau_1) c_{-\Vec{k}l\da}(\tau_2)
        c_{-\Vec{k}'m\da}^\dagger(\tau_3)c_{\Vec{k}'m'\ua}^\dagger(\tau_4)
        \rangle
 \end{eqnarray}
through the equation,
 \begin{eqnarray*}
   P\!=\!\Vec{g}^\dagger \chi \Vec{g}\!
      =\!\!\sum_{ll'mm'}\!\sum_{\omega_n \omega_n'}
   g_{ll'}^\ast(\omega_n)\chi_{ll',mm'}(i\omega_n,i\omega_n')g_{mm'}(\omega_n'). \end{eqnarray*}
Here $\Vec{g}$ is the form factor describing the pairing symmetry 
in orbital-frequency space, which is either even or odd in $\omega$ 
for each orbital component. 
We adopt $g_{\rm odd}(\omega)={\rm sign}(\omega)$ and
$g_{\rm even}(\omega)\equiv 1$ here.

The two-particle Green function (\ref{eq:chi}) is obtained from the 
Bethe-Salpeter equation,
 \begin{align}\label{eq:b-s}
   \chi=\chi_0+\chi_0\Gamma\chi,
 \end{align}
where $\chi$ is defined in eq.(\ref{eq:chi}), 
$\chi_0$ the bare two-particle Green
function calculated as a product of the one-particle Green function, 
$G_{\ua}(\Vec{k},i\omega)G_{\da}(-\Vec{k},-i\omega)$, 
summed over the momenta, 
and $\Gamma$ the vertex function, all of them being 
matrices with respect to orbital and frequency indices.
Within the DMFT we can replace $\Gamma$ with the local one,
 \begin{equation*}
  \Gamma^{\rm loc}=(\chi_0^{\rm loc})^{-1} -(\chi^{\rm loc})^{-1},
 \end{equation*}
in the limit of infinite dimensions\cite{mh89}, where 
$\chi_0^{\rm loc}$ and $\chi^{\rm loc}$ are respectively 
the bare and the dressed 
two-particle Green functions for the effective impurity model, which are 
computed in the QMC.

 \begin{figure}[t]
 \centering{
 \includegraphics[width=7cm,height=5.5cm]{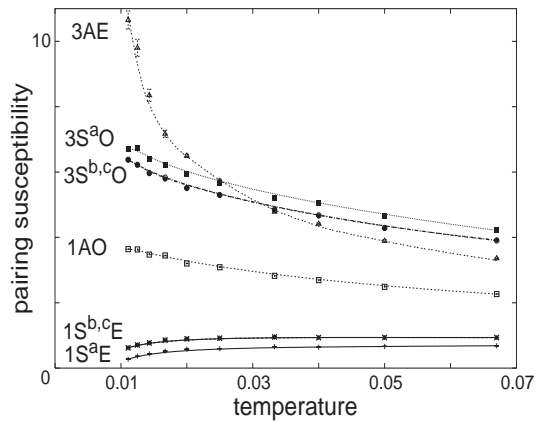}
 \caption{Temperature dependence of the pairing susceptibilities
          for $n\!=\!1.8$, $U'\!=\!0.7$, $J\!=\!0.4$, $U=U'+2J=1.5$.
          The symbols denote pairing symmetries 
          (1: spin-singlet, 3: triplet; 
          $\rm S^{a,b,c}$: orbital-symmetric 
          (eq.(\ref{eq:basis})), A: antisymmetric; 
          E: even-, O: odd-frequency).
          }
 \label{fig:u07j04n18}}
 \end{figure}

 \begin{figure}[t]
 \centering{
 \includegraphics[width=7cm,height=5.5cm]{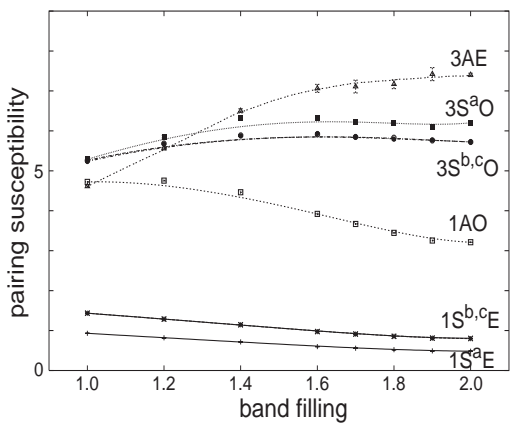}
 \caption{Band-filling dependence of the pairing susceptibilities
          for $\beta\!=\!60$, $U'\!=\!0.7$, $J\!=\!0.4$, $U\!=\!U'+2J\!=\!1.5$ 
          with the same abbreviation for pairing symmetries as in 
          Fig.\ref{fig:u07j04n18}. 
          }
 \label{fig:u07j04b60}}
 \end{figure}

The temperature dependence of the pairing susceptibilities is shown 
in Fig.\ref{fig:u07j04n18} for $n\!=\!1.8$, $U'\!=\!0.7$, $J\!=\!0.4$, and 
$U\!=\!U'\!+\!2J\!=\!1.5$.
We can see that the spin-triplet, orbital-antisymmetric, even frequency 
pairing (denoted as 3AE)
becomes dominant at low temperatures.  
The enhancement of such a pairing should be due to Hund's coupling, which 
tends to align electron spins across different orbitals.  
The result is consistent with 
Han's\cite{han04} for the semi-elliptical density of states.  
While in the work of Han the susceptibility diverges at $T/W\!\sim\! 0.06$ for
$J/W\!=\!0.15, U'/W\!=\!0.45$, we do not detect the divergence up to 
$T/W\!=\!1/180$.  This should be 
because electrons are less correlated in the hypercubic lattice than
in the Bethe lattice (which 
the semi-elliptic density of states would represent), 
since the Gaussian density of states for the hypercubic lattice, 
with high-energy tails, has an effectively larger band width.

Figure \ref{fig:u07j04b60} shows the band-filling dependence of the 
pairing susceptibilities
at $T/W\!=\!1/120$ for $U'\!=\!0.7$, 
$J\!=\!0.4$, and $U\!=\!U'+2J\!=\!1.5$. 
A new finding here is that 
the spin-triplet, orbital-antisymmetric (3AE) pairing induced by Hund's coupling remains dominant in a rather wide range ($n\!=\!1.6\!\sim\!2$) of hole-doping
at this temperature. 
It declines, however, at low fillings ($n\!=\!1\!\sim\!1.4$).
This result may be related to a mechanism of superconductivity 
proposed by Capone\cite{cf02} for multi-orbital systems close to the Mott 
transition, where the electron repulsion $U$ is envisaged to 
assist a pairing even for $s$-waves.  

In summary, we have constructed a new discrete transformation 
(which reduces the sign problem and is applicable to doped bands) 
for the exchange and pair-hopping terms in double-orbit models, 
and implemented this in a DMFT+QMC calculation.  
Superconducting susceptibilities of the $s$-wave pairings calculated 
with this method shows that the spin-triplet, orbital-antisymmetric even
frequency paring, enhanced by Hund's coupling and being dominant 
at half-filling, is robust against hole doping.  
We note that it is difficult to investigate ferromagnetism in this model 
even with the present transformation, since the ferromagnetism 
appears only in very strongly correlated regime\cite{mk98}, where the sign 
problem becomes serious again, so this is a future problem. 

Numerical calculations were partly performed on SR8000 in ISSP, University of Tokyo.


\begin{references}
\bibitem{commentelhole} 
$\langle {\rm sign}\rangle$ is larger in Motome-Imada's method 
at half-filling ($n\!=\!2$) due to the electron-hole symmetry,
but it becomes smaller for $U\!=\!U'\!+\!2J$.
\end{references}
\end{document}